\documentclass[11pt,graphicx]{article}
\usepackage{amssymb,amsmath,amsfonts}
\usepackage{graphicx}
\usepackage{graphics}
\usepackage{eepic,epsfig}
\newcommand{\n}{{\not\hspace{-0.5ex}\nabla}}
\newcommand{\A}{{\not\hspace{-0.8ex}A}}

\begin{document}
\title{Induced fermionic current densities by magnetic flux in higher dimensional cosmic string spacetime}
\author{Eug\^enio R. Bezerra de Mello \thanks{E-mail: emello@fisica.ufpb.br}\\
Departamento de F\'{\i}sica-CCEN\\
Universidade Federal da Para\'{\i}ba\\
58.059-970, J. Pessoa, PB\\
C. Postal 5.008\\
Brazil}
\maketitle       
\begin{abstract}
In this paper we analyse the vacuum fluctuations of quantum fermionic current densities, $\langle j^A(x)\rangle$, induced by a magnetic flux running along the core of the conical space produced by the presence of cosmic string, in a $(1+d)-$dimensional bulk, for $2\leq d\leq 5$. In order to develop this investigation we construct an expression for the fermionic propagator, which can be applied for those different dimensions of the spacetime. Two specific analysis for the induced current densities are presented: $(i)$ for massless fields in three and four dimensions, and $(ii)$ for massive fields in all dimensions considered. In the latter specific values for the parameters which codify the presence of the cosmic string and the fractional part of the ratio of the magnetic flux by the quantum one are considered. Although being a very special situation and somewhat unnatural, the corresponding analysis of induced fermionic current densities under this circumstance may shed light on the qualitative behavior of these quantities in a more general situation.
\\
\\PACS numbers: $98.80.Cq$, $11.10.Gh$, $11.27.+d$  
\vspace{1pc}
\end{abstract}
\maketitle
\section{Introduction}
It is well known that different types of topological defects may have been formed in the early Universe after Planck time by a vacuum phase
transition \cite{Kibble,V-S}. Depending on the topology of the vacuum manifold these are domain walls, strings, monopoles and textures. Among
them, cosmic strings and monopoles seem to be the best candidates to be observed. Cosmic strings are linear topologically stable gravitational defects. The gravitational field produced by a cosmic string may be approximated by a planar angle deficit in the two-dimensional sub-space. The simplest theoretical model describing an idealized cosmic string, i.e., straight and infinitely thin, is given by a delta-type distribution for the energy-momentum tensor along the linear defect. This object can also be described by classical field theory. By coupling the energy-momentum tensor associated with the Maxwell-Higgs system investigated by Nielsen and Olesen in \cite{N-O} with the Einstein equations, Garfinkle \cite{Garfinkle} found static cylindrically symmetric solutions representing vortices, as in flat space-time, and shown that asymptotically the two dimensional spaces perpendicular to the vortices correspond to a plan minus a wedge. Moreover, their core have a non-zero thickness with a non-vanishing magnetic fields inside. In additional publication, Linet \cite{Linet} obtained, as a limiting case, exact solutions for fields equations, corresponding to the spacetime produced by an idealized cosmic string, with the conical parameter being related with the energy per unity length of the vortex. 

Though the recent observation data on the cosmic microwaves background have ruled out cosmic strings as the primary source for primordial density perturbation, they are still candidate for the generation of a number of interesting physical effects such as gamma ray burst \cite{Berezinski}, gravitational waves \cite{Damour} and high energy cosmic ray \cite{Bhattacharjee}. Moreover, in the framework of brane inflation \cite{Sarangi}-\cite{Dvali}, cosmic strings have attracted renewed interest, partly because a variant of their formation mechanism is proposed. 

Although topological defects have been first considered in four-dimensional spacetimes, they have been analysed in higher-dimensional spacetimes in the context of braneworld. By this scenario the defects live in a $n-$dimensions submanifold embedded in a $(4+n)-$dimensional Universe. The cosmic string case, corresponding to a two additional dimensions, has been analysed \cite{Cohen,Ruth}. There it was shown that the gravitational effects of global strings can be responsible for compactification from six to four spacetime dimensions, naturally producing the observed hierarchy between electroweak and gravitational forces. 

The analysis of quantum vacuum fluctuations associated with matter fields in curved spacetime are relevant in a semiclassical theory of gravity \cite{Birrel}. The validity conditions under which the vacuum expectation value (VEV) of the energy-momentum tensor can act as the source for a semiclassical gravitational field are discussed in \cite{Ford,Singh}. In this context, the VEV of the energy-momentum tensor in the idealized cosmic string spacetime, have been considered in \cite{scalar}-\cite{scalar4} and \cite{ferm}-\cite{ferm3}, for scalar and fermionic fields, respectively. It has been shown that these quantities depend on the parameter which codify the conical structure of the two-geometry.  Moreover, considering the presence of magnetic flux running along the cosmic strings, there appears an additional contributions to the corresponding vacuum polarization effects associated with charged fields \cite{charged}-\cite{Spin}.\footnote{Recently the fluxes by gauge fields play an important role in higher dimensional models including braneworld scenario (see for example \cite{Dou}).} Also by imposing boundary conditions on quantum fields, additional shifts in the vacuum expectation values of physical quantities, such as energy density and stress take place. This is the well-known Casimir effects. The analysis of Casimir effects in the idealized cosmic string space-time have been developed for a scalar \cite{Mello}, vector \cite{Mello1} and fermionic fields \cite{Aram1}, obeying boundary conditions on the cylindrical surfaces.\footnote{Also vacuum polarization effects induced by a composite topological defect has been analysed in \cite{Mello3}.}

Additionally to the contribution for the energy-momentum tensor associated with charged quantum fields, the presence of a magnetic flux along the cosmic string induces vacuum current densities, as well. This phenomenon has been investigated for massless scalar field in \cite{Sira} and recently for massive one in \cite{Yu}. There the authors have shown that a vacuum induced azimuthal current densities, $\langle j^\varphi\rangle$, take place if the ratio of the magnetic flux by the quantum one has a non-vanishing fractional part. \footnote{The analysis of electric current densities induced by  magnetic flux have been investigated in \cite{Serebryani}-\cite{Flek} in a flat spacetime.} So in a more general analysis about the influence of vacuum polarization effects on the semiclassical theory of gravity, induced currents produce additional contribution. This quantity is among the most important characteristics of the vacuum state. Though the corresponding operator is local, due to the global nature of the vacuum, the VEV of the current density describes the global properties of the bulk and carry important informations about the structure of the defect core and the magnetic flux. In addition to describe the physical structure of the quantum fields at a given point, the current density acts as source of the electromagnetic field in the Maxwell's equations. It therefore plays an important role in modeling a self-consistent dynamics involving the electromagnetic fields.

Although the analysis of the VEVs of the energy-momentum tensor associated with charged fermionic fields in a cosmic string spacetime having a magnetic flux line running along its core, has been developed by many authors; as far as we know the investigation of the corresponding induced current densities has been restricted for a three-dimensional spacetime only in \cite{BB}. So the objective of the paper is to extend this analysis as general as possible. Adopting this line of investigation, here we shall calculate the vacuum fluctuations of fermionic current densities in a $(1+d)-$dimensional cosmic string spacetime, considering massless and massive fields. As we shall see these current densities depend crucially on the fractional part of the ratio of the magnetic flux by the quantum one. 

This paper is organized as follows: In Section \ref{sec2} we briefly present some important results obtained in \cite{Jean}, associated with fermionic propagators in an arbitrary $(1+d)-$dimensional cosmic string space-time, with $d\geq 2$, in the presence of a magnetic flux. Although the main objective is to investigate induced currents in a higher-dimensional spaces, the formalism obtained allows us to develop similar analysis for a three-dimensional spacetime, as well. Moreover, in this section we also present a closed expression for massive fermionic Green function, expressed in terms of finite sum of Macdonald function, $K_\nu$, for specific values of the parameters which codify the presence of cosmic string and the fractional part of the ratio of the magnetic flux by the quantum one. In Section \ref{sec3} we calculate induced vacuum fermionic current densities for massless and massive fields. For massless fields, we explicitly calculate all components of fermionic current densities for three and four dimensions and shown that only azimuthal components are different from zero. For massive fields and for specific conditions above mentioned, we calculate also all components of fermionic currents for $d=2, \  3, \ 4, \ 5$.  We shown that only for $d=2$ there appears a non-vanishing charge density. In Section \ref{conc}, we summarize the most important results obtained. The Appendix \ref{A1} contains some details related with the analysis of radial and azimuthal current densities associated with massless fields in three-dimensional spacetime. In this paper we shall use $\hbar=G=c=1$, denote the electric charge by $e$ and use for the metric tensor the signature $+2$.

\section {The fermionic propagator}
\label{sec2}
In this section we briefly review the obtainment of the fermionic propagator derived in previous paper \cite{Jean}, in the $(1+d)-$dimensional cosmic string spacetime in the presence of a magnetic flux running along its core. By using the generalized cylindrical coordinate with the cosmic string on the subspace defined by $r=0$, being $r\geq0$ the polar coordinate, the corresponding metric tensor is defined by the line element below:
\begin{eqnarray}
\label{cs0}
	ds^2=g_{MN}dx^Mdx^N=-dt^2+dr^2+\alpha^2r^2d\varphi^2+\sum_{i}(dx^i)^2 \ . 
\end{eqnarray}
The coordinates system reads: $x^M=(t,r,\varphi,x^i)$, with $\varphi\in[0, \ 2\pi]$, and $t, \ x^i\in(-\infty, \ \infty)$. The parameter $\alpha$, smaller than unity, codify the presence of the string and is given, in a four dimensional spacetime, in terms of the linear mass density of the string by $\alpha=1-4\mu$. Specifically for a six-dimensional space, $i=3$, the geometry represented by (\ref{cs0}) corresponds to two dimensional conical surface transverse to a $3$-flat brane.

In order to calculate the fermionic propagator, we shall start the construction of the formalism admitting that the spacetime is six-dimensional. So we shall adopt for the $8\times8$ Dirac matrices, the representation below, constructed in terms of the $4\times 4$ ones \cite{B-D,Moha}:
\begin{eqnarray}
\label{gamma}
\Gamma^0=\left( 
\begin{array}{cc}
0&\gamma^0 \\
\gamma^0&0 
\end{array} \right) \ , \
\Gamma^r=\left( 
\begin{array}{cc}
0&{\hat{r}}\cdot{\vec\gamma} \\
{\hat{r}}\cdot{\vec\gamma}&0 
\end{array} \right) \ , \
\Gamma^\varphi=\frac1{\alpha r}\left( 
\begin{array}{cc}
0&{\hat{\varphi}}\cdot{\vec\gamma} \\
{\hat{\varphi}}\cdot{\vec\gamma}&0 
\end{array} \right) \ , \nonumber\\
\Gamma^{x}=\left( 
\begin{array}{cc}
0&\gamma^{(3)}\\
\gamma^{(3)}&0 
\end{array} \right) \ , \
\Gamma^{y}=\left( 
\begin{array}{cc}
0&i\gamma_5\\
i\gamma_5&0 
\end{array} \right) \ , \
\Gamma^{z}=\left( 
\begin{array}{cc}
0&I \\
-I&0 
\end{array} \right) \ ,
\end{eqnarray}
where  $\gamma_5=i\gamma^0\gamma^1\gamma^2\gamma^3$, $I$ represents the $4\times 4$ identity matrix and ${\hat{r}}$ and ${\hat{\varphi}}$ stand the ordinary unit vectors in cylindrical coordinates. This set of matrices satisfies the Clifford algebra $\{\Gamma^M, \ \Gamma^N  \}=-2g^{MN}I_{(8)}$.

Let us now introduce an infinitesimally thin magnetic flux running along the string. This will be implemented by considering a six-dimensional vector potential below
\begin{eqnarray}
	A_M=A\partial_M\varphi \ ,
\end{eqnarray}
being $A=\frac{\Phi}{2\pi}$. 

The fermionic propagator on curved spacetime and in the presence of a gauge fields, can be expressed in terms of a bispinor, ${\cal{D}}_F(x,x')$, by \cite{Birrel}
\begin{equation}
\label{Sf} 
{\cal{S}}_F(x,x')=\left(i\n+e\A+M\right){\cal{D}}_F(x,x') \ ,
\end{equation} 
where the covariant derivative operator reads
\begin{equation}
\n=\Gamma^M(\partial_M+\Pi_M) \ ,
\end{equation}
being $\Pi_M$ the the spin connection given in terms of the $\Gamma-$matrices
\begin{eqnarray}
\Pi_M=-\frac14\Gamma_N\nabla_M\Gamma^N \ ,
\end{eqnarray}
and
\begin{eqnarray}
\A=\Gamma^MA_M \ .
\end{eqnarray}
Moreover, the bispinor must obey the differential equation below \cite{Spin},
\begin{eqnarray}
\label{D}
\left[{\Box}-ieg^{MN}(D_M A_N)+ie\Sigma^{MN}F_{MN}-2ieg^{MN}A_M\nabla_N\right. \nonumber\\
\left.-e^2g^{MN}A_M A_N-M^2-\frac14{\cal R}\right]{\cal{D}}_F(x,x')&=&-\frac1{\sqrt{-g}}\delta^6(x-x')I_{(8)} \ ,
\end{eqnarray}
with
\begin{equation}
\Sigma^{MN}=\frac14[\Gamma^M,\Gamma^N] \ , \ D_M=\nabla_M-ieA_M \ . 
\end{equation}
$\cal{R}$ represents the scalar curvature and the generalized d'Alembertian is expressed by 
\begin{eqnarray}
\Box=g^{MN}\nabla_M\nabla_N=g^{MN}\left(\partial_M\nabla_N+\Pi_M\nabla_N-\{^S_{MN}\}\nabla_S\right) \ .
\end{eqnarray}

For the system under consideration the differential operator $\cal{K}$, which acts on the left hand side of (\ref{D}), reduces to
\begin{eqnarray}
\label{K0}
{\cal{K}}&=&{\Delta}+\frac i{\alpha^2 r^2}(1-\alpha)\Sigma^3_{(8)}\partial_\varphi-\frac1{4\alpha^2 r^2}(1-\alpha)^2+\frac e{\alpha^2 r^2} (1-\alpha)A\Sigma^3_{(8)}\nonumber\\
&-&\frac{2ie}{\alpha^2 r^2}A\partial_\varphi-\frac{e^2}{\alpha^2 r^2}A^2-M^2\ , 
\end{eqnarray}
where
\begin{equation}
\Sigma^3_{(8)}=\left( \begin{array}{cccc}
  \Sigma^3& 0\\ 
0 & \Sigma^3
                      \end{array}
               \right) \ , \ {\rm with} \
\Sigma^3=\left( \begin{array}{cccc}
  \sigma^3& 0\\ 
0 & \sigma^3
                      \end{array}
               \right) \ 
\end{equation}
and\footnote{For this geometry the only non-vanishing spin connection is $\Pi_\varphi=\frac i2(1-\alpha)\Sigma^3_{(8)}$.}  
\begin{equation}
\label{delta}
{{\Delta}}=-\partial_t^2+\partial_r^2+\frac 1r\partial_r+\frac 1{\alpha^2 r^2}\partial^2_\varphi+\partial_x^2+\partial_y^2+\partial_z^2 \ .
\end{equation}

Because $\Sigma^3_{(8)}$ is a diagonal matrix the bispinor, ${\cal{D}}_F(x,x')$ can be given in terms of a $2\times2$ bispinor, ${\cal{D}}^{(2)}_F(x,x')$, in a diagonal form. The latter obeying the $2\times2$ matrix differential operator below: 
\begin{eqnarray}
\label{K}
\left[{\Delta}+\frac i{\alpha^2 r^2}(1-\alpha)\sigma^3\partial_\varphi-\frac1{4\alpha^2 r^2}(1-\alpha)^2+\frac e{\alpha^2 r^2} (1-\alpha)A\sigma^3\right.\nonumber\\
\left.-\frac{2ie}{\alpha^2 r^2}A\partial_\varphi-\frac{e^2}{\alpha^2 r^2}A^2-M^2\right]{\cal{D}}^{(2)}_F(x,x')= -\frac1{\sqrt{-g}}\delta^6(x-x') I_{(2)} \ .
\end{eqnarray}

On basis of these results, for this six-dimensional cosmic string space-time, the fermionic propagator can be expressed as shown below:
\begin{eqnarray}
\label{SF}	S_F(x,x')=\left[i\Gamma^0\partial_t+i\Gamma^r\partial_r+i\Gamma^\varphi\partial_\varphi+i\Gamma^i\partial_i-\frac{1-\alpha}2\Gamma^\varphi\Sigma^3_{(8)}+\frac {e\Phi}{2\pi}\Gamma^\varphi+M\right]D_F(x,x') \ .
\end{eqnarray}
                                          
The calculation of the induced current can be given in terms of the Euclidean bispinor. The latter been related with the ordinary Feynman propagator  by the relation  ${\cal D}_E(\tau,\vec{r}; \tau', \vec{r'}) = -i {\cal D}_F(x,x')$, with $t=i\tau$ \cite{Birrel}. In the following we shall consider the Euclidean bispinor only. 

In order to calculate the $2\times 2$ Euclidean bispinor, ${\cal{D}}^{(2)}_E(x,x')$, we shall first obtain the complete set of normalized eigenfunctions of the Euclidean version, ${\bar{\cal{K}}}$, of the differential operator given in (\ref{K}):
\begin{eqnarray}
\label{K1}
	{\bar{\cal{K}}}\Phi_\lambda(x)=-\lambda^2\Phi_\lambda(x) \ ,
\end{eqnarray}
with $\lambda^2\geq 0$. In this way ${\cal{D}}^{(2)}_E(x,x')$ is given by
\begin{eqnarray}
\label{D2}
	{\cal{D}}^{(2)}_E(x,x')=\sum_{\lambda^2}\frac{\Phi_\lambda(x)\Phi_\lambda^\dagger(x')}{\lambda^2}=\int_0^\infty \ ds\sum_{\lambda^2} \Phi_\lambda(x)\Phi_\lambda^\dagger(x') \ e^{-s\lambda^2} \ .
\end{eqnarray}
The eigenfunctions of (\ref{K1}) can be specified by a complete set of quantum number associated with operators that commute with ${\bar{\cal{K}}}$ and among themselves. They are: $p_\tau=-i\partial_\tau$, $p_i=-i\partial_i$ for $i=3,\ 4, 5$ , $L_\varphi=-i\partial_\varphi$ and the spin operator, $\sigma_3$. Let us denote these quantum numbers by $(k^\tau, \ k^i, \ n, \ \sigma)$, where $(k^\tau, \ k^i)\in (-\infty, \ \infty)$, $n=0, \ \pm 1, \ \pm 2, \ ... \ $, $\sigma=\pm 1$, and $p$, that satisfies the relation $\lambda^2=p^2+k^2+M^2$, assuming value in the interval $[0, \ \infty)$. Imposing on the eigenfunctions the Dirichlet boundary condition on the string,\footnote{A more general boundary condition implies singular solutions in $r=0$. It is our intention to extend the analisys presented in this paper admitting singular solutions.} 
\begin{eqnarray}
\label{BC}
\Phi_\lambda(t,r=0,\varphi,{\vec{x}})=0 \ , 
\end{eqnarray}
the corresponding eigenfunctions are:
\begin{eqnarray}
\label{Phi}
	\Phi_\lambda^{(+)}(x)&=&\frac{e^{ik.x}{\sqrt{p}}}{[\alpha(2\pi)^{N+1}]^{1/2}}e^{in\varphi}J_{|\nu^+|/\alpha}(pr){\cal{\omega}}^{(+)} \ ,  \nonumber\\
	\Phi_\lambda^{(-)}(x)&=&\frac{e^{ik.x}{\sqrt{p}}}{[\alpha(2\pi)^{N+1}]^{1/2}}e^{in\varphi}J_{|\nu^-|/\alpha}(pr){\cal{\omega}}^{(-)}
\end{eqnarray}
where
	\begin{eqnarray}
	\label{omega}
	{\cal{\omega}}^{(+)}=\left(
\begin{array}{cc}
1 \\
0
\end{array} \right) \  , \ 
	{\cal{\omega}}^{(-)}=\left(
\begin{array}{cc}
0 \\
1
\end{array} \right) \ \ .
\end{eqnarray}
In the above expressions $J_\mu$ represents the Bessel function, and $\nu^{\pm}=n\pm\frac{(1-\alpha)}2-\gamma$, where we have defined $eA=e\frac{\Phi}{2\pi}=\gamma$, being $\gamma$ $\in(0, \ 1)$.

Finally with the help of \cite{Grad} we can express ${\cal{D}}^{(2)}_E(x,x')$ in a integral representation:
\begin{eqnarray}
\label{D2b}
	{\cal{D}}^{(2)}_E(x,x')&=&\frac1{\alpha(4\pi)^{N/2+1}}\int_0^\infty \frac{ds}{s^{N/2+1}}e^{-\frac{(\Delta x)^2+r^2+r'^2}{4s}-M^2s} \ \sum_n e^{in(\varphi-\varphi')}\times\nonumber\\
&&\left( \begin{array}{cc}
I_{|\nu^+|/\alpha}(rr'/2s)&0\\
0& I_{|\nu^-|/\alpha}(rr'/2s))
\end{array} \right)  \ \ ,
\end{eqnarray}
where $I_\mu(z)$ is the modified Bessel function.

Although we have developed this formalism for a six-dimensional cosmic string space-time, we can adapt it for a three, four and five dimensional spaces. The reason resides in the representations adopted for the Dirac matrices. For three dimensional space the corresponding Dirac matrices are given in terms of $2\times 2$ Pauli matrices; and for a four and five dimensions, they are given in terms of the $4\times 4$ off-diagonal matrices of (\ref{gamma}). Moreover, in (\ref{D2b}) we should assume $N=1, \ 2, \ 3, \ 4$, which are related with the dimensions of the space considered.

Unfortunately, it is not possible to integrate over the variable $s$ in (\ref{D2b}) and to express the Euclidean bispinor in a closed form. In fact all the calculations of vacuum polarizations considering massive fields involve very complicated integrals. So in order to present more workable expressions, the massless limit of the corresponding results are adopted. Here in this paper, we shall explicitly calculate the renormalized induced fermionic current for different dimensions of the space-time in closed expressions. So, in order to do that we shall consider two different limiting cases which allow us to reach this objective as shown in the following subsections. 

\subsection{Massless fermionic fields}
In this subsection we provide the bispinor associated with a charged massless fermionic fields in an arbitrary dimension. Taking $M=0$ in (\ref{D2b}) we obtain \cite{Grad},
\begin{eqnarray}
\label{D2c}
	{\cal{D}}^{(2)}_E(x,x')&=&\frac1{\alpha(2\pi)^{\frac{N+3}2}}\frac1{(rr')^{\frac{N}2}}\frac1{(-\sinh u)^{\frac{N-1}2}} \times \nonumber\\
&&\sum_n e^{in(\varphi-\varphi')}\left( 
\begin{array}{cc}
Q_{|\nu^+|/\alpha-1/2}^{\frac{N-1}2}(\cosh u)&0\\
0& Q_{|\nu^-|/\alpha-1/2}^{\frac{N-1}2}(\cosh u)) 
\end{array} \right)  \ \ ,
\end{eqnarray}
where
\begin{eqnarray}
\label{cs}
	\cosh u=\frac{(\Delta x)^2+r^2+r'^2}{2rr'} \ .
\end{eqnarray}
In the above expression, $Q^\lambda_\nu$ is the associated Legendre function. 

For a three dimensional space, $N=1$, the Legendre function can be expressed in a integral form:
\begin{eqnarray}
\label{Q}
Q_{\nu-1/2}(\cosh u)=\frac1{\sqrt{2}}\int_u^\infty dt  \frac{e^{-\nu t}}{\sqrt{\cosh t- \cosh u}} \ .
\end{eqnarray}
Substituting the above expression into (\ref{D2c}), it is possible to obtain a closed expression for the bispinor by developing the sum over the quantum number $n$. Admitting that the parameters $\delta^+=\frac{(1-\alpha)}2-\gamma$ and $\delta^-=\frac{(1-\alpha)}2+\gamma$ that appear in $\nu_\pm$, are both restricted to the interval, $0\leq\delta^\pm<1$, the Euclidean bispinor reads:
\begin{eqnarray}
\label{D3a}
{\cal{D}}_E^{(2)}(x',x)=\frac{1}{4\pi^2\alpha\sqrt{2rr'}}
\int_u^\infty \frac{dt}{\sqrt{\cosh t- \cosh u}}\left( 
\begin{array}{cc}
S^{(+)}(t)&0\\
0&S^{(-)}(t)
\end{array} \right)  \ ,  
\end{eqnarray} 
 where
\begin{eqnarray}
\label{S}
	S^{(\pm)}(t)=\frac{e^{\mp i(\varphi-\varphi')}\sinh(\delta^\pm t/\alpha)-\sinh[(\delta^\pm-1)t/\alpha]}{\cosh(t/\alpha)-\cos(\varphi-\varphi')} \ .
\end{eqnarray}

For a four dimensional space, $N=2$, the Legendre function can be represented by a simpler expression:
\begin{eqnarray}
	Q_{\nu-1/2}^{1/2}(\cosh u)=i{\sqrt\frac\pi2}\frac{e^{-\nu u}}{{\sqrt{\sinh u}}} \ .
\end{eqnarray} 
In this case, substituting this function into (\ref{D2c}) and developing similar procedure as before, we obtain
\begin{eqnarray}
\label{D3b}
{\cal{D}}_E^{(2)}(x',x)=\frac{1}{8\pi^2\alpha rr'\sinh u}
\left( 
\begin{array}{cc}
S^{(+)}(u)&0\\
0&S^{(-)}(u))
\end{array} \right)  \ , 
\end{eqnarray}  
where $S^{(\pm)}(u)$ has the same functional form as $S^{(\pm)}(t)$ in (\ref{S}). \footnote{The expressions for the bispinor in five and six dimensions are given in \cite{Jean}; however, because we only analyse the three and four dimensional case for massless fields, we shall not repeat them here.} 

\subsection{Massive fermionic fields}
\label{Heat}
In the analysis of vacuum polarization effects associated with quantum scalar fields in a cosmic string spacetime, Smith \cite{scalar1}, Davies and Sahni \cite{scalar2}, and Souradeep and Sahni \cite{scalar3}, have shown that when the parameter $\alpha$ is equal to the inverse of an integer number, i.e., when $\alpha=\frac1q$, being $q$ an integer number, the corresponding Green functions can be expressed in terms of $q$ images of the Minkowski spacetime functions. Recently the image method was also used in \cite{Mello} to provide closed expressions for massive scalar Green functions for a higher-dimensional cosmic string space-time. The reason for the application of this method is because the order of the Bessel functions which appear in the derivations of the Green functions becomes an integer number. Unfortunately, for the fermionic case, the order of the Bessel function depends on the factor $\frac{(1-\alpha)}2$ coming from the spin connection. However, if we consider a charged fermionic field in the presence of a magnetic flux running along the string, an additional factor will be present, the ratio of the magnetic flux by the quantum one, $\gamma$. For the case where $\gamma$ is equal to $\frac{(1-\alpha)}2$ and $\alpha=1/q$, the order of the Bessel function becomes an integer number, and in this case it is possible to use the image method to obtain the fermionic propagator in a closed form.\footnote{This special situation was also used in \cite{Aram} to analysis fermionic vacuum polarizations by composite topological defect.} Although being a very special situation and somewhat unnatural, once that by a typical grand unified theory a realistic value for $\alpha$ is very  close the unity, the analysis of induced fermionic current under this circumstance may shed light on the qualitative behavior of this quantity for non-integer $q$. 

In order to derive a more compact expression for the Euclidean bispinor in the above mentioned situation, let us go back to the integrand of (\ref{D2b}), the heat kernel:
\begin{eqnarray}
	{\cal{K}}(x,x';s)=\frac{e^{-\frac{(\Delta x)^2+r^2+r'^2}{4s}-M^2s}}{\alpha(4\pi s)^{N/2+1}}\sum_n e^{in\Delta\varphi}
	\left(\begin{array}{cc}
I_{|\nu^+|/\alpha}(rr'/2s)&0\\
0&I_{|\nu^-|/\alpha}(rr'/2s)
\end{array} \right) \ .
\end{eqnarray}

Defining
\begin{eqnarray}
\label{Ka}
 {\bar{K}}^{(+)}(x,x';s)=\sum_n e^{in\Delta\varphi}I_{|\nu^+|/\alpha}(rr'/2s) \ ,
\end{eqnarray}
with $\nu^+=n+\frac{(1-\alpha)}2-\gamma$, the order of the modified Bessel function becomes an integer number. For this case, with help of \cite{Pru,Jean} we can write:
\begin{eqnarray}
\label{Kaa}
 {\bar{K}}^{(+)}(x,x';s)=\sum_n e^{in\Delta\varphi}I_{nq}(rr'/2s)=\frac1q\sum_{k=0}^{q-1} e^{\frac{rr'}{2s} \cos\left(\frac{\Delta\varphi}q+\frac{2\pi k}q\right)} \ .
\end{eqnarray}
Analogously, defining 
\begin{eqnarray}
\label{Kb}
 {\bar{K}}^{(-)}(x,x';s)=\sum_n e^{in\Delta\varphi}I_{|\nu^-|/\alpha}(rr'/2s)=e^{i\Delta\varphi}\sum_n e^{in\Delta\varphi}I_{nq+1}(rr'/2s) \ ,
\end{eqnarray}
after some steps we obtain
\begin{eqnarray}
\label{Kab}
 {\bar{K}}^{(-)}(x,x';s)=e^{i(1-1/q)\Delta\varphi}\ \frac1q\sum_{k=0}^{q-1} e^{\frac{rr'}{2s} \cos\left(\frac{\Delta\varphi}q+\frac{2\pi k}q\right)} e^{-\frac{2ik\pi}{q}} \ .	
\end{eqnarray}
Finally substituting ${\bar{K}}^{(+)}$ and ${\bar{K}}^{(-)}$ into the heat kernel and integrating on the variable $s$ we find with the help of \cite{Grad},
\begin{eqnarray}
\label{Dq}
		{\cal{D}}_E^{(2)}(x',x)=\frac{M^{N/2}}{(2\pi)^{N/2+1}}\sum_{k=0}^{q-1}\frac1{(\rho_k)^{N/2}}K_{N/2}(M\rho_k)	\left(\begin{array}{cc}
1&0\\
0&e^{i(1-1/q)\Delta\varphi}e^{-2i\pi k/q}
\end{array} \right) \ ,
\end{eqnarray}
with 
\begin{eqnarray}
\label{rho}
	\rho^2_k=(\Delta x)^2+r^2+r'^2-2rr'\cos\left(\Delta\varphi/q+2\pi k/q\right) \ .
\end{eqnarray}
 
Now having derived the basic expressions to calculate the fermionic propagator, we leave for the next section the main objective of our paper, that is the calculations of induced fermionic current densities. 

\section{Induced current densities}
\label{sec3}
The fermionic current density operator is an expression bilinear on the field. By using the general definition for fermionic propagator,
\begin{eqnarray}
	i{\cal{S}}_F(x,x')=\langle0|T(\Psi(x){\bar{\Psi}}(x'))|0\rangle \ ,
\end{eqnarray}
with ${\bar{\Psi}}=\Psi^\dagger\Gamma^0$, we can express the VEV of fermionic current density as the trace of the product of the Dirac matrix by the fermionic propagator shown below:
\begin{eqnarray}
\label{current}
	\langle j^A(x)\rangle=-ie\lim_{x'\to x}{\rm Tr}(\Gamma^A S_F(x',x)) \  .
\end{eqnarray}

In the two next subsections, we shall present the steps developed to obtain vacuum expectation values of the fermionic current densities for massless and massive fields.

\subsection{Massless field}
In this subsection we shall calculate the induced fermionic current densities for a massless fields, considering three and four dimensional cosmic string spacetimes.

\subsubsection{Three dimensional case}
\label{sec311}
From the general expression for fermionic propagator given by (\ref{SF}), for a three-dimensional spacetime, it is express in terms of a $2\times2$ bispinor as shown below:
\begin{eqnarray}
S_F(x',x)=i\left[i\sigma^3\partial_t-\sigma^r\partial_r-\frac{1}{\alpha r}\sigma^\varphi\partial_\varphi-\frac{i}{\alpha r}\frac{(1-\alpha)}{2} \sigma^\varphi\sigma^3+\frac{i}{\alpha r}\gamma\sigma^\varphi\right]{\cal{D}}_E^{(2)}(x',x) \ .
\end{eqnarray}

Let us start the calculation by the time component of the current density. Because the bispinor above is in a diagonal form, in the calculation of charge density, $\langle j^0\rangle$, only the time derivative term survives when we take the trace over the matrices product $\sigma^3S_F(x',x)$. So the corresponding VEV is given by
\begin{eqnarray}
	\langle j^0(x)\rangle=ie\lim_{x'\to x}{\rm Tr}(\partial_\tau{\cal{D}}_E^{(2)}(x',x) ) \  .
\end{eqnarray}

Here we shall write the bispinor in the form,
\begin{eqnarray}
\label{DI}
{\cal{D}}_E^{(2)}(x',x)=\frac{1}{4\pi^2\alpha\sqrt{2rr'}}
\left( 
\begin{array}{cc}
I^{(+)}(u)&0\\
0&I^{(-)}(u)
\end{array} \right)  \ . 
\end{eqnarray} 
The relevant calculation here is
\begin{equation}
	\partial_\tau I^{(\pm)}(u)=\partial_\tau\int_u^\infty\frac{dt}{\sqrt{\cosh t-\cosh u}}S^{(\pm)}(t) \ .
\end{equation}
Let us first take $\Delta r=\Delta\varphi=0$. Introducing a new variable $t:=2 \ {\rm arcsinh(y/\sqrt{2})}$, the function $I^{(\pm)}(u)$ reads:
\begin{eqnarray}
	I^{(\pm)}(z)=2\int_{\sqrt{z-1}}^\infty \frac{dy}{\sqrt{2+y^2}}\frac{S^{(\pm)}(2 \ {\rm arcsinh(y/\sqrt{2})})}{\sqrt{y^2-(z-1)}} 
\end{eqnarray}
with $z=\cosh u=1+\frac{\Delta\tau^2}{2r^2}$.

In order to calculate the VEV of the charge density operator, we shall write the above function as the sum of two terms:
\begin{eqnarray}
	I^{(\pm)}(z)&=&I_1^{(\pm)}(z)+I_2^{(\pm)}(z)=2\int_{\sqrt{z-1}}^1 \frac{dy}{\sqrt{2+y^2}}\frac{{\tilde{S}}^{(\pm)}(y)}{\sqrt{y^2-(z-1)}}\nonumber\\
	&+&2\int_1^\infty \frac{dy}{\sqrt{2+y^2}}\frac{{\tilde{S}}^{(\pm)}(y)}{\sqrt{y^2-(z-1)}} \ ,
\end{eqnarray}
where ${\tilde{S}}^{(\pm)}(y)=S^{(\pm)}(t(y))$. Subtracting and adding into the integrand of $I_1^{(\pm)}$ the first term of the expansion 
\begin{eqnarray}
	\frac{{\tilde{S}}^{(\pm)}(y)}{\sqrt{2+y^2}}\approx \frac\alpha y+O(y)
\end{eqnarray}
we get
\begin{eqnarray}
	I_1^{(\pm)}=I_1^{(\pm)fin}+I_1^{(\pm)sing} \ ,
\end{eqnarray}
where
\begin{eqnarray}
I_1^{(\pm)fin}(z)=2\int_{\sqrt{z-1}}^1\frac{dy}{\sqrt{y^2-(z-1)}}\left(\frac{{\tilde{S}}^{(\pm)}(y)}{\sqrt{2+y^2}}-\frac\alpha y \right)
\end{eqnarray}
and
\begin{eqnarray}
I_1^{(\pm)sing}(z)=2\int_{\sqrt{z-1}}^1\frac{dy}{\sqrt{y^2-(z-1)}}\frac\alpha y =\frac{2\alpha\eta}{\sqrt{z-1}}  \ ,
\end{eqnarray}
being $\eta={\rm arctg}\left(\sqrt{\frac{2-z}{z-1}}\right)$. In the limit $\Delta\tau\to0$, the above expression behaves as
\begin{eqnarray}
I_1^{(\pm)sing}(z)\approx\frac{\sqrt{2}r\alpha\pi}{\Delta\tau}+O(1) \ .
\end{eqnarray}
Substituting this expression into (\ref{DI}), the singular part gives,
\begin{eqnarray}
	{\cal{D}}_E^{(0)}(x',x)=\frac1{4\pi\Delta\tau}I_{(2)} \ ,
\end{eqnarray}
which coincides with the Euclidean bispinor for a flat three-dimensional space-time. This is the only term that contains divergence in the calculation of the charge density. To obtain a finite result, we should extract its divergent part. This can be done in a manifest form by subtracting from the complete bispinor, the standard Euclidean Green function given above. So the renormalized bispinor reads:
\begin{eqnarray}
{\cal{D}}_E^{(2)}(x',x)_{Ren}=\frac{1}{4\pi^2\alpha r\sqrt{2}}
\left( 
\begin{array}{cc}
I_1^{(+)fin}(z)+I_2^{(+)}(z)&0\\
0&I_1^{(-)fin}(z)+I_2^{(-)}(z)
\end{array} \right)  \ . 
\end{eqnarray}  
Defining a new variable $b=\sqrt{z-1}$, we can write:
\begin{eqnarray}
I_1^{(\pm)fin}(z)=\int_b^1\frac{dy}{\sqrt{y^2-b^2}}f^{(\pm)}(y) \ \ {\rm with} \ \ f^{(\pm)}(y)= 2\left(\frac{{\tilde{S}}^{(\pm)}(y)}{\sqrt{2+y^2}}-\frac\alpha y \right) \ .
\end{eqnarray}
So, 
\begin{eqnarray}
\partial_\tau I_1^{(\pm)fin}(z)=\frac1{r\sqrt{2}}\frac{d}{db}\int_b^1\frac{dy}{\sqrt{y^2-b^2}}f^{(\pm)}(y)   \ .
\end{eqnarray}
Because the integrand is divergent at the point $y=b$, we have to change the variable $y\to by$ before applying the Leibnitz formula for derivative of an integral. After taken the first order derivative, defined $x=by$ and developed some intermediate steps, we find:
\begin{eqnarray}
\frac{d}{db}\int_b^1\frac{dy}{\sqrt{y^2-b^2}}f^{(\pm)}(y)=b\left[\int_b^1\frac{dx}{\sqrt{x^2-b^2}}\frac{d}{dx}\left(\frac{f^{(\pm)}(x)}{x}\right)
-\frac{f^{(\pm)}(1)}{\sqrt{1-b^2}}\right] \ .
\end{eqnarray}
Now we can take the limit $b\to 0$. The term inside the bracket gets a finite result, consequently the complete expression vanishes in this limit. The analysis of the time derivative of $I_2^{(\pm)}(z)$ in the limit $\Delta\tau\to0$ is very simple and a vanishing result is promptly obtained. 

As conclusion we can affirm that there is no charge density induced by the magnetic flux for the physical situation investigated.

As to the radial current, $<j^r>$, two distinct contributions are relevant in the product, $\sigma^rS_F(x',x)$. They are expressed as shown below:
\begin{eqnarray}
{\rm Tr}\left(\gamma^r S_F(x',x)\right)={\rm Tr}\left\{\left[\partial_r+\frac1{\alpha r}\left(i\sigma^3\partial_\varphi- \frac{1}{2}(1-\alpha)+\gamma\sigma^3\right)\right]{\cal{D}}_E^{(2)}(x',x)\right\} \ .
\end{eqnarray}
First we analyse the contribution due to the radial derivative. It is divergent in the coincidence limit, and this divergence coincides with the corresponding one given by the standard Green function. Renormalizing this contribution in the same way as we did before we get a finite and non-vanishing result.\footnote{In order to make a short discussion, some details involved on this calculations are similar to that will appear in the next one.} The second contribution is due to the other three terms. The latter is finite at the coincidence limit. Taking both contributions together we obtain the imaginary expression bellow:
\begin{eqnarray}
\label{jr3d}
\langle j^r(x)\rangle_{Ren.}=\frac{ie}{16\pi^2r^2}\int_0^\infty \frac{dv}{\sinh(v/2)}\frac1{\cosh(v/\alpha)-1}G(v,\alpha,\gamma) \ ,
\end{eqnarray}
with
\begin{eqnarray}
G(v,\alpha,\gamma)&=&\frac{4\cosh(v/2\alpha)\sinh(v/2)\cosh(v\gamma/\alpha)}{\alpha^2}+\frac{2\sinh(v)}{\cosh(v)-1}(1-\cosh(v/\alpha))\nonumber\\
&-&\frac{8\gamma\sinh(v/2\alpha)\sinh(v/2)\sinh(v\gamma/\alpha)}{\alpha^2} \ .
\end{eqnarray}
In appendix \ref{A1} it is proved that (\ref{jr3d}) vanishes for $\gamma$ smaller than $\frac12$, which is compatible with the condition imposed for the parameter $\delta^\pm$ in (\ref{S}). In fact, because we are implicitly assuming that $\alpha<1$, the condition on the parameter $\delta^+=(1-\alpha)/2-\gamma$, imposes that $1/2-\gamma$ must be bigger than zero. Consequently our analysis is valid only for $0\leq\gamma<1/2$; moreover, the condition $\gamma<1/2$, naturally appears under specific situation obeyed by the parameters $\alpha$ and $\gamma$ given in the beginning of subsection \ref{Heat}.

In the calculation of azimuthal component of the current density we have:
\begin{eqnarray}
\label{azimu}
{\rm Tr}\left(\gamma^\varphi S_F(x',x)\right)=-\frac{i}{\alpha r}{\rm Tr}\left\{\left[\sigma^3\partial_r+\frac{i}{\alpha r}\left(\partial_\varphi+ \frac{i}{2}(1-\alpha)\sigma^3-i\gamma\right)\right]{\cal{D}}_E^{(2)}(x',x)\right\} \ .
\end{eqnarray}
The first contribution due to the radial derivative is given by:
\begin{eqnarray}
\label{rad}
\partial_r{\cal{D}}_E^{(2)}(x',x)&=&\frac1{4\pi^2\alpha}\left[\frac{\partial}{\partial r}\left(\frac1{\sqrt{2rr'}}\right)\right]
	\left(\begin{array}{cc}
I^{(+)}(u)&0\\
0&I^{(-)}(u)
\end{array} \right)\nonumber\\
&+&\frac1{4\pi^2\alpha}\frac1{\sqrt{2rr'}}\frac{\partial}{\partial r}\left(\begin{array}{cc}
I^{(+)}(u)&0\\
0&I^{(-)}(u)
\end{array} \right) ,
\end{eqnarray}
where
\begin{eqnarray}
	I^{(\pm)}(u)=\int_u^\infty\frac{dt}{\sqrt{\cosh t-\cosh u}}S^{(\pm)}(t) \ .
\end{eqnarray}  
At this point we may taken $\Delta\tau=0$, providing $\cosh u=\frac{r'^2+r^2}{2r'r}$, and also $\Delta\varphi=0$.

In the first contribution of (\ref{rad}) we get a divergent result from the terms inside the matrix when we take the coincidence limit. The later  provides
\begin{eqnarray}
	I^{(\pm)}(u)\rightarrow\frac1{\sqrt{2}}\int_0^\infty\frac{dt}{\sinh(t/2)}S^{(\pm)}(t) \ .
\end{eqnarray}
In order to obtain a finite result, we should renormalize the bispinor by subtracting from it its corresponding Euclidean flat bispinor,
\begin{eqnarray}
	{\bar{\cal{D}}}_E^{(0)}(x',x)=\frac1{4\pi\Delta r}I_{(2)} \ .
\end{eqnarray}
Expressing the above Green function in a integral form, we get, for the renormalized contribution, the following expression:
\begin{eqnarray}
\label{rad1}
-\frac1{16\pi^2r^2}\int_0^\infty\frac{dt}{\sinh(t/2)}\left(\begin{array}{cc}
\frac{S^{(+)}(t)}\alpha-S^{(0)}(t)&0\\
0&\frac{S^{(-)}(t)}\alpha-S^{(0)}(t)
\end{array} \right) \ ,
\end{eqnarray}
where $S^{(0)}(t)$ is obtained from (\ref{S}) by taking $\delta^{\pm}=0$. 

The second contribution of (\ref{rad}) comes from $\frac{\partial I^{(\pm)}(u)}{\partial r}$. Using a procedure similar to the adopted in the calculation of the charge density, we can prove that $\partial_rI_1^{(\pm)fin}(u)$ and $\partial_rI_2^{(\pm)}(u)$ go to zero at the coincidence limit. So, the only effective non-vanishing contribution is given by (\ref{rad1}).

As to the contributions due to the azimuthal derivative, spin connection and magnetic flux in (\ref{azimu}), they provide a finite results in the coincidence limit, so it is not required any renormalization procedure. Finally in the coincidence limit we have:
\begin{eqnarray}
\left(\partial_\varphi+ \frac{i}{2}(1-\alpha)\sigma^3-i\gamma\right){\cal{D}}_E^{(2)}(x',x)\longmapsto&&\frac{i}{8\pi^2\alpha r}\int_0^\infty  \frac{dt}{\sinh(t/2)}\frac1{\cosh(t/\alpha)-1}\times\nonumber\\
&&\left(\begin{array}{cc}
{\tilde{S}}^{(+)}(t)&0\\
0&{\tilde{S}}^{(-)}(t)
\end{array} \right) \ ,
\end{eqnarray}
where
\begin{eqnarray}
{\tilde{S}}^{(\pm)}(t)=\pm\delta^{\pm}[\sinh(\delta^\pm t/\alpha)+\sinh[(1-\delta^\pm)t/\alpha]]\mp\sinh(\delta^\pm t/\alpha) \ .
\end{eqnarray}

Now let us now write down the complete result for the azimuthal current. It reads
\begin{eqnarray}
\label{jaz3d}
\langle j^\varphi(x)\rangle_{Ren.}=\frac{e}{8\pi^2\alpha^3r^3}\int_0^\infty\frac{dt}{\sinh(t/2)}\frac1{\sinh^2(t/2\alpha)}F(t,\alpha,\gamma) \ ,
\end{eqnarray}
where
\begin{eqnarray}
F(t,\alpha,\gamma)=\cosh(t/2)\left[\cosh(t/2\alpha)\sinh(t\gamma/\alpha)-2\gamma\sinh(t/2\alpha)\cosh(t\gamma/\alpha)\right] \ .
\end{eqnarray}

\begin{figure}[tbph]
\begin{center}
\begin{tabular}{cc}
\epsfig{figure=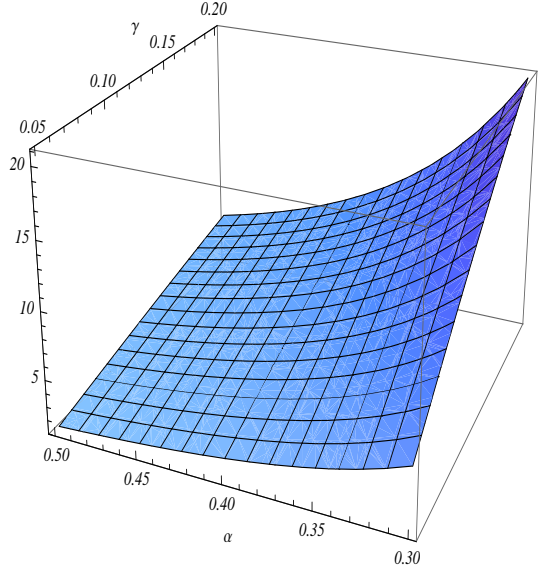, width=7.5cm, height=7.5cm,angle=0} 
\end{tabular}
\end{center}
\caption{This graph provides the behaviors $\frac{8\pi^2r^3}{e} \langle j^\varphi(x)\rangle_{Ren.}$ as functions of $\alpha$ and $\gamma$, admitting these parameters varying in the interval $[0.30, \ 0.50]$ and $[0.05, \ 0.20]$, respectively.}
\label{fig}
\end{figure}

By the above expression we can easily see that $F(t,\alpha,\gamma)$ vanishes for $\gamma=0$. The behavior of the the azimuthal current as function of the parameters $\alpha$ and $\gamma$ can only be provided numerically. So in figure \ref{fig} we exhibit the behavior of $\frac{8\pi^2r^3}{e} \langle j^\varphi(x)\rangle_{Ren.}$ as function of $\alpha$ and $\gamma$. \footnote{Because the expression found for (\ref{S}) has been obtained admitting that $0<\delta^\pm<1$ our numerical analysis is according to this conditions}. 

The appearance of the above induced current is a direct consequence of the well known Aharanov-Bohm effect. In principle it should be a periodic function of $\gamma$; however, because the integrand of (\ref{jaz3d}) has been obtained under specific conditions, for a fixed value of $0<\alpha<1$ the dependence of $<j^\varphi(x)>_{Ren}$ with the parameter $\gamma$ is positive and convex downwards, presenting a maximum value at some specific point. (This fact is also observed by numerical analysis.)

\subsubsection{Four dimensional case}
For a four dimensional spacetime, the Dirac matrices in the cosmic string spacetime, can be given by
\begin{eqnarray}
\Gamma^0=\gamma^{(0)}, \ \Gamma^r={\hat{r}}\cdot{\vec{\gamma}} \ , \ \Gamma^\varphi=\frac1{\alpha r}{\hat{\varphi}}\cdot{\vec{\gamma}} \ {\rm and} \ \Gamma^z=\gamma^{(3)} \ .
\end{eqnarray}
The corresponding fermionic propagator can be obtained by (\ref{SF}), being the bispinor the $4\times 4$ diagonal matrix below
\begin{eqnarray}
\label{D4a}
{\cal{D}}_E(x',x)=
\left( 
\begin{array}{cc}
{\cal{D}}_E^{(2)}(x',x)&0\\
0&{\cal{D}}_E^{(2)}(x',x)
\end{array} \right)  \ ,
\end{eqnarray}
${\cal{D}}_E^{(2)}(x',x)$ being given by
\begin{eqnarray}
{\cal{D}}_E^{(2)}(x',x)=\frac{1}{8\pi^2\alpha rr'\sinh u}
\left( 
\begin{array}{cc}
S^{(+)}(u)&0\\
0&S^{(-)}(u))
\end{array} \right)  \ , 
\end{eqnarray}  
with
\begin{eqnarray}
	S^{(\pm)}(u)=\frac{e^{\mp i(\varphi-\varphi')}\sinh(\delta^\pm u/\alpha)-\sinh[(\delta^\pm-1)u/\alpha]}{\cosh(u/\alpha)-\cos(\varphi-\varphi')} \ ,
\end{eqnarray}
\begin{eqnarray}
\delta^\pm=\frac{(1-\alpha)}2\mp\gamma \ \ {\rm and} \ \ \cosh u=\frac{\Delta\tau^2+\Delta z^2+r^2+r'^2}{2rr'} \ .
\end{eqnarray}

Because the bispinor is expressed in terms o elementary functions, the calculations for all components of the induced current are relatively simple. Below we present the main steps of the calculations.

Let us first analyse zero component of the current. For the charge density the only non-vanishing trace contribution is given in term of time derivative:
\begin{equation}
{\rm Tr}(\gamma^0S_F(x',x))=-{\rm Tr}(\partial_\tau {\cal{D}}_E(x',x)) \ .
\end{equation}
Here we can take first $\Delta\varphi=\Delta z=\Delta r=0 $, consequently in the coincidence limit,
\begin{eqnarray}
\partial_\tau\frac1{\sinh u}S^{(\pm)}(u)\longmapsto-\frac{4\alpha r^2}{(\tau'-\tau)^3}+O((\tau'-\tau)) \ .
\end{eqnarray} 
This divergence can be removed by subtracting from the bispinor above the standard Green function for a flat four dimensional spacetime. As consequence we obtain a vanishing result for the charge density.

For the radial current we have:
\begin{equation}
\label{rad-curr}
{\rm Tr}(\gamma^rS_F(x',x))=i{\rm Tr}\left\{\left[-i\partial_r+\frac1{\alpha r}\left(\Sigma^{(3)}\partial_\varphi+\frac{i}{2}(1-\alpha)- i\gamma\Sigma^{(3)}\right)\right] {\cal{D}}_E(x',x)\right\} \ .
\end{equation}
The first contribution is due to the radial derivative. We may take $\Delta\varphi=\Delta z=\Delta\tau=0$, so in the coincidence limit we have,
\begin{eqnarray}
\partial_r\frac1{rr'}S^{(\pm)}(u)\longmapsto-\frac{4\alpha}{(r-r')^3}-\frac{1-\alpha^2+6\delta^{(\pm)}(\delta^{(\pm)}-1)}{6\alpha r^2}+O(r-r') \ .
\end{eqnarray}
Substituting back this expression into the bispinor, the first term on the right hand side above, coincides with the corresponding one given by the standard Green function. Again applying the renormalization procedure adopted in the previous calculations, a finite contribution is left. It is given below:
\begin{eqnarray}
\lim_{r'\to r}\partial_r{\cal{D}}_E^{(2)}(x',x)_{Ren.}=-\frac1{48\pi^2\alpha^2r^3}\left( 
\begin{array}{cc}
1-\alpha^2+6\delta^{+}(\delta^{+}-1)&0\\
0&1-\alpha^2+6\delta^{-}(\delta^{-}-1)
\end{array} \right)  \ . 
\end{eqnarray} 
As to the second contribution we can analyse the operator below,
\begin{eqnarray}
&&\left(\sigma^3\partial_\varphi+\frac i2(1-\alpha)-i\gamma\sigma^3\right){\cal{D}}_E^{(2)}(x',x)\longmapsto\frac i{24\pi^2\alpha^2r^2}\times\nonumber\\
&&\left( 
\begin{array}{cc}
\delta^{+}[1+\delta^{+}(2\delta^{+}-3)]&0\\
0&\delta^{-}[1+\delta^{-}(2\delta^{-}-3)]
\end{array} \right)  \ . 
\end{eqnarray}
Now considering both contributions above together and taking the trace, we get a vanishing result for the radial current.

As to the azimuthal current, the contributions which provide a non-vanishing traces are given by:
\begin{equation}
\label{azi-curr}
{\rm Tr}(\gamma^\varphi S_F(x',x))=-i{\rm Tr}\left\{\left[\Sigma^3\partial_r+\frac1{\alpha r}\left(i\partial_\varphi-\frac{1}{2}(1-\alpha)\Sigma^3+ \gamma\right)\right] {\cal{D}}_E(x',x)\right\} \ .
\end{equation}

Comparing the above expression with the similar one obtained for the radial current, we can see an important difference between them: The presence of the overall matrix $-i\Sigma^3$ multiplying the left hand side of (\ref{rad-curr}). Because the steps to be followed are formally the same as the previous one, we shall not repeat them. Taking the trace of all contributions, we obtain a real and non-vanishing value to the azimuthal current:
\begin{equation}
\label{azi-curr-0}
\langle j^\varphi(x)\rangle_{Ren.}=\frac{e\gamma}{12\pi^2(\alpha r)^4}(1+3\alpha^2-4\gamma^4) \ .
\end{equation}
Because the expressions found for (\ref{D3b}) has been obtained admitting that $0\leq\delta^\pm<1$, the above result can only be valid for $0\leq\gamma<1$.

Finally as to the current along the $z-$direction, we have:
\begin{eqnarray}
{\rm Tr}(\gamma^3 S_F(x',x))={\rm Tr}\left(\partial_z {\cal{D}}_E(x',x)\right) \ .
\end{eqnarray}
Taking here $\Delta\varphi=\Delta r=\Delta\tau=0$, we obtain $\cosh u=1+\frac{\Delta z^2}{2r^2}$, consequently in the coincidence limit we may write:
\begin{eqnarray}
\partial_z\frac1{\sinh u}S^{(\pm)}(u)\longmapsto-\frac{4\alpha r^2}{(z-z')^3}+O((z-z')) \ .
\end{eqnarray}
The first term on the right hand side of the above expression coincides with similar one obtained from the standard Green function. So by applying on this calculation the same renormalization procedure that we have applied until now, we get a vanishing result for $\langle j^z(x)\rangle_{Ren.}$

\subsection{Massive case}
Here we shall analyse the VEV of fermionic current densities associated with massive fields in high dimensional cosmic string space times under the circumstance specified in \ref{Heat}. In order to do that we shall use the bispinor (\ref{Dq}) to calculate the fermionic propagator. However, before to start the calculation, it is interesting to analyse this bispinor in the coincidence limit. We can verify that taking $x'\to x$ this bispinor is divergent, and that this divergence comes exclusively from the $k=0$ component. So, in order to obtain a finite and well defined results for the currents, we should apply some renormalization prescription. As in the previous calculation, we shall apply this procedure in a manifest form, by subtracting from (\ref{Dq}) its $k=0$ component. So the renormalized bispinor is express as,
\begin{eqnarray}
\label{Dren}
		{\cal{D}}^{(2)}(x',x)_{Ren.}=\frac{M^{N/2}}{(2\pi)^{N/2+1}}\sum_{k=1}^{q-1}\frac1{(\rho_k)^{N/2}}K_{N/2}(M\rho_k)	\left(\begin{array}{cc}
1&0\\
0&e^{i(1-1/q)\Delta\varphi}e^{-2i\pi k/q}
\end{array} \right) \ .
\end{eqnarray}

Adopting $N=1, \ 2, \ , 3, \ 4$ we can construct the fermionic propagator for spacetimes of three, four, five and six dimensions.

\subsubsection{Three dimensional case}
\label{321}
The renormalized fermionic propagator in a conical three dimensional spacetime, can be expressed by:
\begin{eqnarray}
S_F(x',x)_{Ren.}=i\left[i\sigma^3\partial_t-\sigma^r\partial_r-\frac{1}{\alpha r}\sigma^\varphi\partial_\varphi-\frac{i}{\alpha r}\frac{(1-\alpha)}{2} \sigma^\varphi\sigma^3+\frac{i}{\alpha r}\gamma\sigma^\varphi+M\right]{\cal{D}}_E^{(2)}(x',x)_{Ren.} \ ,
\end{eqnarray}
being
\begin{eqnarray}
		{\cal{D}}^{(2)}(x',x)_{Ren.}=\frac{M^{1/2}}{(2\pi)^{3/2}}\sum_{k=1}^{q-1}\frac1{\sqrt{\rho_k}}K_{1/2}(M\rho_k)	\left(\begin{array}{cc}
1&0\\
0&e^{i(1-1/q)\Delta\varphi}e^{-2i\pi k/q}
\end{array} \right) \ ,
\end{eqnarray}
with
\begin{eqnarray}
	\rho^2_k=(\Delta\tau)^2+r^2+r'^2-2rr'\cos\left(\Delta\varphi/q+2\pi k/q\right) \ .
\end{eqnarray}

The first calculation to be realized is for the charge density. We may take, $\Delta\varphi=\Delta r=0$ in the above expressions. The only non-vanishing trace of matrix, $\gamma^0S_F(x',x)$, are given below:
\begin{eqnarray}
\label{j03m}
{\rm Tr}(\gamma^0S_F(x',x))_{Ren.}=i{\rm Tr}\{[i\partial_\tau+M\sigma^3]{\cal{D}}^{(2)}(x',x)_{Ren.}\} \ .
\end{eqnarray}
Let us first to analyse the time derivative of the bispinor. In the coincidence limit, we promptly obtain:
\begin{eqnarray}
\lim_{\tau'\to\tau}\partial_\tau\frac1{\sqrt{\rho_k}}K_{1/2}(M\rho_k)=0 \ .
\end{eqnarray}
The second contribution is due to the mass term. In the coincidence limit, after some intermediate steps, we find:
\begin{eqnarray}
\label{MS}
{\rm Tr}\left[iM\sigma^3{\cal{D}}^{(2)}(x',x)_{Ren.}\right]=\frac{iM}{4\pi r}\sum_{k=1}^{q-1}\sin(\pi k/q) \ e^{-2Mr\sin(\pi k/q)} \ .
\end{eqnarray}
Substituting this expression into (\ref{j03m}), and using the definition for the VEV of current density (\ref{current}), we obtain:
\begin{eqnarray}
\langle j^0(x)\rangle_{Ren}=\frac{eM}{4\pi r}\sum_{k=1}^{q-1}\sin(\pi k/q) \ e^{-2Mr\sin(\pi k/q)} \ , 
\end{eqnarray}
which presents an exponentially suppressed behavior.

Now we can calculate the induced electric charge by integration as shown below:
\begin{eqnarray}
Q=\int\int dr d\varphi{\sqrt{g_{\varphi\varphi}}}\langle j^0(x)\rangle=\frac{e}{4q}(q-1)=\frac e2\gamma \ . 
\end{eqnarray} 

The next calculation is the radial current. The non-vanishing contributions to this current are given below:
\begin{eqnarray}
{\rm Tr}\left(\gamma^r S_F(x',x)\right)={\rm Tr}\left\{\left[\partial_r+\frac {iq}{ r}\left(\sigma^3\partial_\varphi+ \frac{i}{2}(1-\alpha)-i\gamma\sigma^3\right)\right]{\cal{D}}_E^{(2)}(x',x)\right\} \ .
\end{eqnarray}

For the specific case under consideration, two differential operators naturally emerge. They are:
\begin{eqnarray}
D^+=\partial_r+\frac{iq}r(\partial_\varphi+i\delta^+)=\partial_r+\frac{iq}r\partial_\varphi
\end{eqnarray}
because $\delta^+=0$, and
\begin{eqnarray}
D^-=\partial_r-\frac{iq}r(\partial_\varphi-i\delta^-)
\end{eqnarray}
with $\delta^-=\frac{q-1}{q}$. 

The above operators act on the each diagonal components of the bispinor as follows:
\begin{eqnarray}
\label{D+}
D^+\left[\frac1{\sqrt{\rho_k}}K_{1/2}(M\rho_k)\right]=\partial_r\left[\frac1{\sqrt{\rho_k}}K_{1/2}(M\rho_k)\right]+ \frac{iq}r\partial_\varphi\left[\frac1{\sqrt{\rho_k}}K_{1/2}(M\rho_k)\right] 
\end{eqnarray}
and
\begin{eqnarray}
\label{D-}
&&D^-\left[\frac1{\sqrt{\rho_k}}K_{1/2}(M\rho_k)e^{i(1-1/q)\Delta\varphi}e^{-2\pi k/q}\right]=\partial_r\left[\frac1{\sqrt{\rho_k}}K_{1/2} (M\rho_k)e^{i(1-1/q)\Delta\varphi}e^{-2\pi k/q}\right]\nonumber\\
&-& \frac{iq}r\partial_\varphi\left[\frac1{\sqrt{\rho_k}}K_{1/2}(M\rho_k)e^{-2\pi k/q}\right]
-\frac qr\delta^-\left[\frac1{\sqrt{\rho_k}}K_{1/2} (M\rho_k)e^{i(1-1/q)\Delta\varphi}e^{-2\pi k/q}\right] \ .
\end{eqnarray}

Because the dependences of the functions inside the brackets, on the radial polar coordinate and angular variable, are trough $\rho_k$, we may use the identity:
\begin{eqnarray}
\partial_r\equiv\sin(\pi k/q)\frac{d}{d\rho_k} \ \ {\rm and} \ \ \partial_\varphi\equiv\frac qr\cos(\pi k/q)\frac{d}{d\rho_k} \ .
\end{eqnarray}
Doing this, we can obtain simplified results for (\ref{D+}) and (\ref{D-}):
\begin{eqnarray}
\label{DD}
D^+\left[\frac1{\sqrt{\rho_k}}K_{1/2}(M\rho_k)\right]&=&ie^{-i\pi k/q}\frac{d}{d\rho_k}\left[\frac1{\sqrt{\rho_k}}K_{1/2}(M\rho_k)\right]\nonumber\\
D^-\left[\frac1{\sqrt{\rho_k}}K_{1/2}(M\rho_k)e^{i(1-1/q)\Delta\varphi}e^{-2\pi k/q}\right]&=&-ie^{-i\pi k/q}\frac{d}{d\rho_k} \left[\frac1{\sqrt{\rho_k}}K_{1/2}(M\rho_k)\right] \ .
\end{eqnarray}
Now returning to the calculation of the radial current density, we have to take the trace of the resulting matrix. That means that we have to add both results above, consequently we find $\langle j^r\rangle_{Ren}=0$.

For the azimuthal current, the relevant terms to be analysed are:
\begin{eqnarray}
{\rm Tr}\left(\gamma^\varphi S_F(x',x)\right)=-\frac{iq}{r}{\rm Tr}\left\{\left[\partial_r+\frac {iq}{ r}\left(\sigma^3\partial_\varphi+ \frac{i}{2}(1-\alpha)-i\gamma\sigma^3\right)\right]\sigma^3{\cal{D}}_E^{(2)}(x',x)\right\} \ .
\end{eqnarray}
In this calculation we can adopt the same approach that we have used in previous calculation; however, here there exist an important difference when compared with the later. The matrix $\sigma^3$ appears multiplying the bispinor. For this case, repeating the same steps as we did before, we get a non-vanishing result when take the trace of final result. We obtain:
\begin{eqnarray}
\lim_{x'\to x}{\rm Tr}\left(\gamma^\varphi S_F(x',x)\right)=\frac{iq}{8\pi r^3}\sum_{k=1}^{q-1}\frac{1+2Mr\sin(\pi k/q)}{\sin(\pi k/q)}e^{-2Mr\sin(\pi k/q)} \ .
\end{eqnarray}
Finally the current density reads,
\begin{eqnarray}
\langle j^\varphi(x)\rangle_{Ren}=\frac{eq}{8\pi r^3}\sum_{k=1}^{q-1}\frac{1+2Mr\sin(\pi k/q)}{\sin(\pi k/q)} e^{-2Mr\sin(\pi k/q)} \ .
\end{eqnarray}

We may analyse this expression in two limiting cases:
\begin{itemize}
\item In the limit $Mr>>\frac1{\sinh(\pi k/q)}$, the main contributions come from $k=1$ and $k=q-1$ and the leading term is
\begin{eqnarray}
\langle j^\varphi(x)\rangle_{Ren}\approx\frac{eqM}{2\pi r^2} e^{-2Mr\sin(\pi/q)} \ .
\end{eqnarray}
This component presents an exponential suppressed behavior.
\item Massless limit,
\begin{eqnarray}
\label{jaz3dq}
\langle j^\varphi(x)\rangle_{Ren}=\frac{eq}{8\pi r^3} \sum_{k=1}^{q-1}\frac{1}{\sin(\pi k/q)} \ .
\end{eqnarray}
In appendix \ref{A1} it is shown that the general expression for the the azimuthal current density in three-dimensions for massless fermionic fields, Eq. (\ref{jaz3d}), reproduces the result above when we substitute in that expression $\alpha=1/q$ and $\gamma=(q-1)/2q$. 
\end{itemize}

\subsubsection{Four dimensional case} 
\label{d4}
For a four dimensional spacetime, the renormalized $2\times 2$ bispinor reads,
\begin{eqnarray}
{\cal{D}}^{(2)}(x',x)_{Ren.}=\frac{M}{4\pi^2}\sum_{k=1}^{q-1}\frac1{\rho_k}K_1(M\rho_k)	\left(\begin{array}{cc}
1&0\\
0&e^{i(1-1/q)\Delta\varphi}e^{-2i\pi k/q}
\end{array} \right) \ ,
\end{eqnarray}
with
\begin{eqnarray}
	\rho^2_k=\Delta\tau^2+\Delta z^2+r^2+r'^2-2rr'\cos\left(\Delta\varphi/q+2\pi k/q\right) \ .
\end{eqnarray}
On the other hand the fermionic propagator is,
\begin{eqnarray}	S_F(x,x')=i\left[i\gamma^0\partial_t+i\gamma^r\partial_r+\frac{iq}{r}\gamma^\varphi\partial_\varphi+i\gamma^3\partial_z-\frac{q-1}{2r}\gamma^\varphi\Sigma^3_{(8)}+\frac {q\gamma}{r}\gamma^\varphi+M\right]{\cal{D}}_E(x,x') \ ,
\end{eqnarray}
being ${\cal{D}}_E(x,x')={\rm diag}({\cal{D}}^{(2)}(x',x), \ {\cal{D}}^{(2)}(x',x))$.

For the calculation of charge density, we have,
\begin{eqnarray}
{\rm Tr}(\gamma^0S_F(x',x))_{Ren.}=i{\rm Tr}\{[i\partial_\tau+M\gamma^0]{\cal{D}}(x',x)_{Ren.}\} \ .
\end{eqnarray} 
Taking the time derivative of the bispinor and then the coincidence limit, we get a zero result. As to the last term in above expression we get a vanishing result because the trace of the product $\gamma^0{\cal{D}}(x',x)_{Ren.}$ is zero. So as final result we have $\langle j^0\rangle=0$.

For the radial current, we can also apply the same procedure as we did in the three dimensional case. The reason is because we can reduce the present analysis for a $2\times 2$ matrix formalism. Moreover, because we did not have to explicitly calculate the derivative on the Macdonald functions in (\ref{DD}), we can infer by the same reasons as before, that $\langle j^r\rangle=0$.

For the azimuthal current density we have:
\begin{eqnarray}
{\rm Tr}(\Gamma^\varphi S_F(x',x))_{Ren.}=-\frac{iq}{r}{\rm Tr}\left\{\left[\partial_r+ \frac{iq}{r}\left(\Sigma^3\partial_\varphi+\frac{i(q-1)}{2q}-i\gamma\Sigma^3\right)\right]\Sigma^3{\cal{D}}_E(x,x')\right\} \ .
\end{eqnarray} 
Once more, applying to the derivative of the radial and azimuthal coordinates of the bispinor the same procedure used before, we finally obtain, after some intermediate steps, the following result for the azimuthal current density:
\begin{eqnarray}
\langle j^\varphi(x)\rangle=-\frac{eqM}{2\pi^2 r^2}\sum_{k=1}^{q-1}\left[MK_0(2Mr\sin(\pi k/q))-\frac1{r\sin(\pi k/q)}K_1(2Mr\sin(\pi k/q))\right] \ .
\end{eqnarray}
Two limiting cases deserve to be analyzed:
\begin{itemize}
\item In the limit $Mr>>\frac1{\sinh(\pi k/q)}$, 
\begin{eqnarray}
\langle j^\varphi(x)\rangle\approx-\frac{eqM^2}{2\pi^2 r^2}\sqrt{\frac\pi{Mr\sin(\pi/q)}} e^{-2Mr\sin(\pi/q)} \ .
\end{eqnarray}
\item Massless limit,
\begin{eqnarray}
\label{j4d}
\langle j^\varphi(x)\rangle=\frac{eq}{4\pi^2 r^4} \sum_{k=1}^{q-1}\frac{1}{\sin^2(\pi k/q)} \ .
\end{eqnarray}
The summation on the right hand side of (\ref{j4d}) can be written in a closed form. Defining the sum
\begin{eqnarray}
	I_n(x)=\sum_{l=1}^{q-1}\frac1{\sin^n(x+l\pi/q)} \ ,
\end{eqnarray}
for even $n$, $I_n(x)$ can be developed by using the formulas \cite{Mello} 
\begin{eqnarray}
\label{Ir}
	I_{n+2}(x)=\frac{I''_n(x)+n^2I_n(x)}{n(n+1)} \ {\rm being} \ I_2(x)=\frac{q^2}{\sin^2(qx)}-\frac1{\sin^2(x)} \ .
\end{eqnarray}
Because $I_2(0)=\frac{(q^2-1)}3$, we have
\begin{eqnarray}
\label{azi-curr-M}
\langle j^\varphi(x)\rangle=\frac{eq}{12\pi^2 r^4}(q^2-1) \ .
\end{eqnarray}
The above result is an analytical function of $q$, and by analytical continuation is valid for all arbitrary values of $q$.\footnote{Substituting $\alpha=\frac1q$ and $\gamma=\frac{q-1}{2q}$ into (\ref{azi-curr-0}) we obtain (\ref{azi-curr-M}).}
\end{itemize}

For the calculation of current along the $z-$direction we have:
\begin{eqnarray}
{\rm Tr}(\gamma^3S_F(x',x))={\rm Tr}\left(\partial_z{\cal{D}}_E(x,x')\right) \ .
\end{eqnarray}
Taking the derivative of the bispinor and the coincide limit we get a vanishing result.

\subsubsection{Higher-dimensional spacetimes}
The main objective of this section is to calculate fermionic current densities in a five and six-dimensional cosmic string spacetime.

In a generalized $(1+d)-$dimensional idealized cosmic string spacetime the non vanishing singular curvature is located on the $(d-2)-$brane, i.e., the scalar curvature has support only on the $r=0$: $R=4\pi(\alpha^{-1}-1)\delta^{(2)}({\vec{r}})$. For $d\geq4$ this correspond to a $(d-2)-$hypersurface. Considering a magnetic flux along the brane, which has also support on the brane, an azimuthal induced current on the transverse two-dimensional conical surface defined by the coordinates $r$ and $\varphi$ according to (\ref{cs0}), takes place. By using the Maxwell equation, a magnetic field is also induced outside the brane, being directed along it.\footnote{It our intention to examine the behavior of the induced magnetic field, and its consequence as source of the Einstein equations in a backreaction formalism.} Specifically for a six-dimensional bulk, our world can be represented by a flat four-dimensional spacetime having a non-vanishing magnetic field in it. In this context the analysis which will be developed in this subsection, may be useful in a modeling a self-consistent dynamics involving charged quantum fields.

a an azimuthal electric current will be induced on the transverse two-dimensional cone. 

For a five dimensional case, the Dirac matrices are $4\times 4$ ones and can be represented by,
\begin{eqnarray}
\Gamma^0=\gamma^{(0)}, \ \Gamma^r={\hat{r}}\cdot{\vec{\gamma}} \ , \ \Gamma^\varphi=\frac1{\alpha r}{\hat{\varphi}}\cdot{\vec{\gamma}} \  , \ \Gamma^x=\gamma^{(3)} \ {\rm and} \ \Gamma^y=i\gamma_5 \  .
\end{eqnarray}
As to the bispinor, it can be written in a diagonal form in terms of the $2\times2$ bispinor below,
\begin{eqnarray}
		{\cal{D}}^{(2)}(x',x)_{Ren.}=\frac{M^{3/2}}{4\pi^2\sqrt{2\pi}}\sum_{k=1}^{q-1}\frac1{\rho_k^{3/2}}K_{3/2}(M\rho_k)	\left(\begin{array}{cc}
1&0\\
0&e^{i(1-1/q)\Delta\varphi}e^{-2i\pi k/q}
\end{array} \right) \ ,
\end{eqnarray}
with
\begin{eqnarray}
	\rho^2_k=\Delta\tau^2+\Delta x^2+\Delta y^2+r^2+r'^2-2rr'\cos\left(\Delta\varphi/q+2\pi k/q\right) \ .
\end{eqnarray}

The procedure to calculate all components of the fermionic current density here, is very similar with the corresponding one for a four-dimensional space-time analyzed in subsection \ref{d4}. To calculate the charge density we have
\begin{eqnarray}
\label{cd-5}
{\rm Tr}(\gamma^0S_F(x',x))_{Ren.}=i{\rm Tr}\{[i\partial_\tau+M\gamma^0]{\cal{D}}(x',x)_{Ren.}\} \ .
\end{eqnarray} 
Taking first the time derivative of the $k$-component of the bispinor, we obtain
\begin{eqnarray}
\partial_\tau\left[\frac1{\rho_k^{3/2}}K_{3/2}(M\rho_k)\right]=\frac{\Delta\tau}{\rho_k}\frac d{d\rho_k}\left[\frac1{\rho_k^{3/2}}K_{3/2}(M\rho_k)\right] \ ,
\end{eqnarray}
which goes to zero in the coincidence limit. As to the second contribution of (\ref{cd-5}), the matrix $\gamma^0{\cal{D}}(x',x)_{Ren.}$ is traceless. So we conclude that $<j^0(x)>=0$.

For the radial component of the current we need to calculate ${\rm Tr}(\gamma^rS(x,x'))$. For this case the effective operator which acts on the bispinor is
\begin{eqnarray}
\label{op4}
-i\partial_r+\frac1{\alpha r}\Sigma^{(3)}\partial_\varphi+\frac{i}{\alpha r}\frac{(1-\alpha)}2-\frac{i\gamma}{\alpha r}\Sigma^{(3)} \ .
\end{eqnarray}
The contribution due to the extra fifth-coordinate on the Dirac operator is given by $\gamma^r\gamma_5\partial_y$, which provides an off-diagonal matrix. So we conclude that there is no essential difference in the calculation of the radial current density when compared with the corresponding calculation in four dimension. Here we also obtain  $<j^r(x)>=0$.

The components $<j^x(x)>$ and $<j^y(x)>$ are also both equal to zero. The reasons for these facts are similar what happened in the calculation of charge density. The derivative with respect to the coordinates $x$ and $y$ on the the bispinor, provide terms proportional to $\Delta x$ and $\Delta y$, respectively, which go to zero in the coincidence limit, and the products of $\gamma^x$ and $\gamma^y$ by ${\cal{D}}(x',x)_{Ren.}$ are traceless. 

Finally we can see, by using the Dirac operator, and after some intermediate steps, that the only non-vanishing component for the fermionic current density is the azimuthal one:
\begin{eqnarray}
\langle j^\varphi(x)\rangle=\frac{eq}{32\pi^2r^5}\sum_{k=1}^{q-1}\frac{e^{-2Mr\sin(\pi k/q)}}{\sin^3(\pi k/q)}\left(3+6Mr\sin(\pi k/q) + 4M^2r^2\sin^2(\pi k/q)\right) \ .
\end{eqnarray}

Below we present two limiting cases for the above current:
\begin{itemize}
\item Large mass limit,
\begin{eqnarray}
\langle j^\varphi(x)\rangle\approx\frac{eqM^2}{4\pi^2 r^3} \frac{e^{-2Mr\sin(\pi/q)}}{\sin(\pi/q)} \ .
\end{eqnarray}
Here also there appears an exponential suppressed behavior.
\item Massless limit,
\begin{eqnarray}
\langle j^\varphi(x)\rangle=\frac{3eq}{32\pi^2 r^5} \sum_{k=1}^{q-1}\frac{1}{\sin^3(\pi k/q)} \ .
\end{eqnarray}
\end{itemize}

For a six dimensional spacetime, the Dirac matrices have already been defined in (\ref{gamma}). The bispinor is also a $8\times 8$ diagonal matrix:
\begin{eqnarray}
{\cal{D}}_E(x',x)=
\left( 
\begin{array}{cccc}
{\cal{D}}_E^{(2)}(x',x)&0&0&0\\
0&{\cal{D}}_E^{(2)}(x',x)&0&0\\
0&0&{\cal{D}}_E^{(2)}(x',x)&0\\
0&0&0&{\cal{D}}_E^{(2)}(x',x)
\end{array} \right)  \ ,
\end{eqnarray} 
being
\begin{eqnarray}
		{\cal{D}}^{(2)}(x',x)_{Ren.}=\frac{M^{2}}{8\pi^3}\sum_{k=1}^{q-1}\frac1{(\rho_k)^2}K_{2}(M\rho_k)	\left(\begin{array}{cc}
1&0\\
0&e^{i(1-1/q)\Delta\varphi}e^{-2i\pi k/q}
\end{array} \right) \ ,
\end{eqnarray}
with
\begin{eqnarray}
\rho^2_k=\Delta\tau^2+\Delta x^2+\Delta y^2++\Delta z^2+r^2+r'^2-2rr'\cos\left(\Delta\varphi/q+2\pi k/q\right) \ .
\end{eqnarray}

The calculations of all components of the fermionic current in this spacetime will be briefly presented in the follows. For the charge density we have,
\begin{eqnarray}
{\rm Tr}\left(\Gamma^0S_F(x,x')\right)_{Ren}=i{\rm Tr}\{\left[i\partial_\tau+M\Gamma^0\right]{\cal{D}}(x',x)_{Ren.}\} \ .
\end{eqnarray}
The time derivative of the $k$-component of the bispinor goes to zero in the coincidence limit. Moreover, the matrix $\Gamma^0{\cal{D}}(x',x)_{Ren.}$ is traceless. So we conclude that $<j^0(x)>=0$.

For the radial component we need to calculate ${\rm Tr}\left(\Gamma^rS_F(x,x')\right)$. The relevant operator is
\begin{eqnarray}
-i\partial_r+\frac1{\alpha r}\Sigma^3_{(8)}\partial_\varphi+\frac{i}{\alpha r}\frac{(1-\alpha)}{2}-\frac{i\gamma}{\alpha r}\Sigma^3_{(8)} \ .
\end{eqnarray}
This operator is a $8\times 8$ matrix version of the (\ref{op4}). Consequently, by considering previous analysis, we can conclude that $<j^r(x)>=0$.

The components $<j^i(x)>$, for $i=3, \ 4, \ 5$, given in terms of ${\rm Tr}\left(\Gamma^iS_F(x,x')_{Ren.}\right)$, provide vanishing results. The reasons are exactly the same as before: the contributions due to the derivative with respect to the coordinates $x^i$ on the bispinor go to zero at the coincidence limit and the others, given by the product $\Gamma^i\Gamma^A{\cal{D}}(x',x)_{Ren.}$, for $A\neq i$, and $\Gamma^i{\cal{D}}(x',x)_{Ren.}$ are traceless matrices. 

Finally we can see that only the azimuthal component of the fermionic current is different from zero. After some steps we find:
\begin{eqnarray}
\langle j^\varphi(x)\rangle&=&\frac{eqM}{4\pi^3r^5}\sum_{k=1}^{q-1}\frac1{\sin^3(\pi k/q)}\left[2K_1(2Mr\sin(\pi k/q))+2Mr\sin(\pi k/q) K_0(2Mr\sin(\pi k/q)\right.\nonumber\\
&+&\left. M^2r^2\sin^2(\pi k/q))K_1(2Mr\sin(\pi k/q))\right] \ .
\end{eqnarray}

Finally we present below the two limiting cases:
\begin{itemize}
\item Large mass limit,
\begin{eqnarray}
\langle j^\varphi(x)\rangle\approx\frac{eqM^{5/2}}{4\pi^{5/2}r^{7/2}} \frac{e^{-2Mr\sin(\pi/q)}}{\sin^{3/2}(\pi/q)} \ .
\end{eqnarray}
\item Massless limit,
\begin{eqnarray}
\langle j^\varphi(x)\rangle=\frac{eq}{4\pi^3r^6}\sum_{k=1}^{q-1}\frac1{\sin^4(\pi k/q)} \ .
\end{eqnarray}
The above summation can be expressed in terms of an analytical function of $q$, by applying the recurrence relation given in (\ref{Ir}) taken $n=2$ and the expression for $I_2(x)$. Although $I_4(x)$ is a long expression, we only need its value for zero argument. It is $I_4(0)=\frac{(q^2-1)(q^2+11)}{45}$, consequently,
\begin{eqnarray}
\langle j^\varphi(x)\rangle=\frac{eq}{180\pi^3r^6}(q^2-1)(q^2+11) \ .
\end{eqnarray}
\end{itemize}

\section{Conclusion and Discussions}
\label{conc}
In this paper we have investigated the vacuum expectation value of fermionic current densities in a $(1+d)-$dimensional cosmic string spacetimes, for $2\leq d\leq5$, induced by the presence of a magnetic flux line running along its core. In this analysis two limiting cases have been considered. The case where the fermionic fields have no mass, has been considered in three and four dimensions only. There we have shown that only the azimuthal components of the current are different from zero. Specifically for three dimension, we have founded an integral representation for it. In order to provide a better understanding about this current we presented its numerical behavior as function of the parameters which codify the presence of the cosmic string and the ratio of the magnetic flux by the quantum one. For a four-dimensional spacetime we could present the azimuthal component of the current density in a closed form. Considering the particular situation where the parameter $\alpha$ is the inverse of an integer number, and the ratio of the magnetic flux by the quantum one, $\gamma$, is equal to $\frac{(1-\alpha)}{2}$, massive fermionic fields were considered in all dimensions above specified. For this case fermionic propagators could be express in terms of finite sum of modified Bessel function, $K_\mu$, and we were able to provide all non-vanishing components of the induced fermionic current densities in closed forms. The corresponding results were analyzed in the massless limit and large mass limit as well. In the latter a strong exponential decay takes place.

The induced fermionic current densities calculated in this paper satisfy the conservation condition, $\nabla_A \langle j^A(x)\rangle_{Ren.}=0$. Another important point which should be emphasized, is the fact that they depend crucially on the fractional part of the ratio of the magnetic flux by the quantum one. This is a direct consequence of the well known Aharanov-Bohm effect. 

An interesting result which deserves to be mentioned is about the calculation of charge density in three dimensions. As we have seen, for massless fields there is no induced charge density; however, considering massive fields, a non-vanishing density is obtained and an induced electric charge appears. This charge is a fractional part of $e$. 

In odd dimensional spacetime, the Clifford algebra has two inequivalent irreducible representations which cannot be mapped to each other by similarity transformations. They may be represented by:\footnote{See paper by Jeon-Hyuck Park in the site http://conf.kias.re.kr/~brane/2003/gamma.pdf.}
\begin{eqnarray}
\gamma^\mu=(\gamma^1, \ \gamma^2, \ ... , \ \gamma^{d+1}) \ {\rm and} \ \gamma^\mu=(\gamma^1, \ \gamma^2, \ ... , \ -\gamma^{d+1}) \ ,
\end{eqnarray}
where $\gamma^{d+1}=\pm\gamma^{12...d}$ or $\gamma^{d+1}=\pm i\gamma^{12...d}$, depending on the dimensions of the spacetime, being $\gamma^{12..d}$ the product of $\gamma^i$. 

In this present paper we have used $\gamma^{1+d}=\gamma^0$. If we had assumed the second representation for the Dirac matrices, the expression found for the bispinor remains the same; however, there will appear a modification in the fermion propagator given in (\ref{SF}) due to the presence of $\gamma^0$ matrix. This new expression for $S_F(x,x')$, would provide a change in the sign of the charge density and induced electric charge for massive fermionic field case analysed in three dimensions.

Finally we want to say that the renormalized procedure adopted to calculate the VEVs of fermionic current densities, was the point-splitting one. In this way we have investigated the behavior of the fermionic propagators in the coincidence limit and extract all the divergences in a manifest form by using the Haddamard functions.\footnote{Because the cosmic string spacetime is locally flat the Hadamard function coincides with the usual Green function.} 

\section*{Acknowledgment}

The author thanks Conselho Nacional de Desenvolvimento Cient\'\i fico e Tecnol\'ogico (CNPq.) for partial financial support, and also FAPES-ES/CNPq. (PRONEX).

\appendix
\section{Special analisys of induced fermionic current in three dimensional cosmic string spacetime}
\label{A1}
In this appendix we present explicitly the steps developed to prove that the radial current in three dimensions for massless fermionic fields, given by Eq. (\ref{jr3d}), in fact is zero; moreover we show that the azimuthal current density, given for general situation by an integral representation in Eq. (\ref{jaz3d}), coincides with the (\ref{jaz3dq}), when we substitute in the former the planar angle deficit and magnetic flux by specific values.

\subsection{Radial current}
In order to prove that the radial fermionic current for massless fields in three dimensions is zero, we write down the result of the indefinite integral which appears in (\ref{jr3d}). It reads,
\begin{eqnarray}
\int\frac{dv}{\sinh(v/2)}\frac1{\cosh(v/\alpha)-1}G(v,\alpha,\gamma)=\frac4{\sinh(v/2)}-\frac4\alpha\frac{\cosh(v\gamma/\alpha)}{\sinh(v/2\alpha)} \ .
\end{eqnarray}
We can see that the behavior of the integral goes to zero when $v\to 0$ and also goes to zero when $v\to\infty$ for $\gamma$ is smaller than $\frac12$, that is compatible with the condition assumed for the parameter $\delta^\pm$ in (\ref{S}), as we have already discussed in subsection \ref{sec311}.
\subsection{Azimuthal current}
The general expression for the azimuthal current density associated with massless fermionic field in three dimensions can only be provided by an integral representation. In this subsection we shall present that if we consider the special values for the planar angle deficit and magnetic flux exhibited in subsection \ref{Heat}, the integral can be exactly solved and the result coincides with that one found in subsection \ref{321}. 

First let us write (\ref{jaz3d}) adopting $\alpha=1/q$ and $\gamma=(q-1)/2q$. The obtained expression reads:
\begin{eqnarray}
\label{japp}
\langle j^\varphi(x)\rangle_{Ren}=\frac{eq^3}{8\pi^2r^3}\int_{-\infty}^\infty  \frac{dx}{\sinh(x)} \ f(x) \ ,
\end{eqnarray}
with
\begin{eqnarray}
\label{fz}
f(x)=\cosh(x)\sinh(x)\left(1-\coth^2(qx)\right)+\frac{\cosh^2(x)}q\left(\coth(qx)-\tanh(x)\right) \ .
\end{eqnarray}
In this new expression we have introduced the variable $v=2x$ and extend the interval of integration to cover all the real axis $x$. 

We can write $f(x)$ as the sum of two other functions,
\begin{eqnarray}
f(x)=f_1(x)+f_2(x) \ ,
\end{eqnarray}
with
\begin{eqnarray}
f_1(x)&=&\cosh(x)\sinh(x)\left(1-\coth^2(qx)\right)+\frac{\coth(x)}{q^2} \nonumber\\
f_2(x)&=&\frac{\cosh^2(x)}q\left(\coth(qx)-\tanh(x)\right)-\frac{\coth(x)}{q^2} \ .
\end{eqnarray}

The integral in (\ref{japp}),
\begin{eqnarray}
\int_{-\infty}^\infty \frac{dx}{\sinh(x)} \ f(x)  \ ,
\end{eqnarray}
can be evaluated by using calculus of residues. 

Let us investigate the contour integral in the complex $z-$plane of the analytical continuation of the function $f(x)$:
\begin{eqnarray}
\label{int-cont}
\oint \frac{dz}{\sinh(z)} \ f(z) \ .
\end{eqnarray}
The contour to be considered is exhibit in figure $2$: 
\begin{figure}[tbph]
\begin{center}
\begin{tabular}{cc}
\epsfig{figure=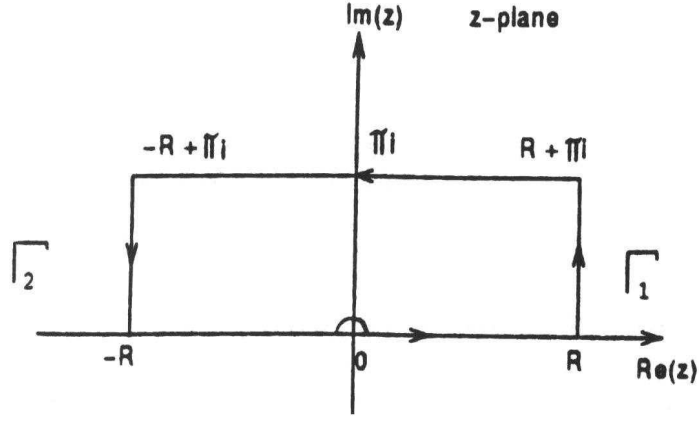, width=9.5cm, height=7.5cm,angle=0} 
\end{tabular}
\end{center}
\caption{The contour used to calculate (\ref{int-cont}).}
\end{figure}

The poles of $f(z)/\sinh(z)$ on the complex $z-$plane are in the imaginary axis, and are given by $z_k=k\pi i/q$, for $k=1, \ 2, \ ... , \ q-1$. 

By the residue theorem we have
\begin{eqnarray}
\oint \frac{dz}{\sinh(z)} \ f(z)=2\pi i\sum{a_{-1}}^{(k)} \ .
\end{eqnarray}
The above integral can be evaluated by using the functions $f_1(z)$ and $f_2(z)$ separately. Let us start with the integral for $f_1(z)$. Because the integrals along the two vertical lines go to zero as $R$ goes to infinity and for the upper horizontal line $f_1(z)/\sinh(z)\to-f_1(z)/\sinh(z)$ we get:
\begin{eqnarray}
\int_{-\infty}^\infty \frac{dz}{\sinh(z)} \ f_1(z)=\pi i\sum{a_{-1}}^{(k)} \ .
\end{eqnarray}
The function $f_1(z)/\sinh(z)$ presents poles of order $2$. By using standard procedure to calculate the residue \cite{Arfken}, we get:
\begin{eqnarray}
a_{-1}^{(k)}=-\frac{i}{q^2}\sin(k\pi/q) \ .
\end{eqnarray}

As to the integral involving $f_2(z)$, we also can write
\begin{eqnarray}
\int_{-\infty}^\infty \frac{dz}{\sinh(z)} \ f_2(z)=\pi i\sum{a_{-1}}^{(k)} \ .
\end{eqnarray}
The poles of the integrand are simple ones and for this case we found
\begin{eqnarray}
a_{-1}^{(k)}=-\frac{i\cos^2(k\pi/q)}{q^2\sin(k\pi/q)} \ .
\end{eqnarray}

So taking into account the both results above for the residues of the integrals, we can see that
\begin{eqnarray}
\int_{-\infty}^\infty \frac{dz}{\sinh(z)} \ f(z)=\frac\pi{q^2}\sum_{k=1}^{q-1}\frac1{\sin(k\pi/q)}  \ .
\end{eqnarray}
Substituting this result into (\ref{japp}) we re obtain (\ref{jaz3dq}).

\end{document}